\documentclass[%
 amsmath,amssymb,
 aps,
]{revtex4-2}

\usepackage{graphicx}% Include figure files
\usepackage{bm}% bold math
\usepackage{float}% table position

\begin{document}

\title{Measurement of changes in muscle viscoelasticity during static stretching using stress-relaxation data}

\author{Yo Kobayashi}
\email{yo.kobayashi@es.osaka-u.ac.jp}%
\affiliation{Graduate School of Engineering Science, Osaka University, Osaka, Japan}%

\author{Daiki Matsuyama}
\affiliation{Graduate School of Engineering Science, Osaka University, Osaka, Japan}%

% \date{\today}

\begin{abstract}

This study investigates how the viscoelasticity of the muscle changes during static stretching by measuring the state of the muscle during stretching using continuous time-series data. We used a device that applied a force to the muscle during stretching and measured the reaction force. The device was attached to the participants, and time-series data of the reaction force (stress-relaxation data) during stretching were obtained. A model using fractional calculus (spring-pot model) was selected as the viscoelastic model for the muscle, in which the data for stress relaxation were fitted on a straight line on a both logarithmic plot. The experimental stress-relaxation results showed that viscoelasticity tended to change abruptly at a particular time during static stretching because the stress-relaxation data were represented by a broken line comprising two segments on the both logarithmic plot. Considering two states of viscoelasticity, before and after the change, the stress-relaxation curve was fitted to the spring-pot model with high accuracy using segment regression (R2 = 0.99). We compared the parameters of the spring-pot model before and after the change in muscle viscoelasticity. By examining these continuous time-series data, we also investigated the time taken for the effects of stretching to become apparent. Furthermore, by measuring the changes in muscle viscoelasticity during static stretching before and after a short-term exercise load of running on a treadmill, we examined the effects of short-term exercise load on the changes in viscoelasticity during static stretching. 


\end{abstract}

\maketitle

\section{\label{sec:introduction}Introduction}

With the recent increase in health and fitness awareness, stretching is considered an important factor for improving health and athletic performance. For athletes and fitness-conscious individuals, stretching the muscles after exercise is necessary to relieve fatigue and prepare them for the next workout. In addition, stretching is important to prevent muscle soreness and injuries that can occur after exercise. Stretching after exercise promotes blood flow and the flexibility of stiff muscles. Maintaining flexible muscles reduces strain on muscles and joints and prevents muscle and joint injuries. In modern society, the working population is aging and the physical burden on workers is increasing. Stretching is also important for reducing the physical burden on workers and in preventative medicine. Thus, the importance of stretching is multifaceted in today's increasingly health-conscious society. Therefore, a detailed investigation on how stretching changes muscles that have been stiffened by exercise is important.

Related research describes the changes in muscle properties due to stretching and exercise and changes in muscle condition during stretching. First, previous studies on the changes in muscle properties induced by stretching are explained. Magnusson et al. investigated passive torque and electromyographic activity with static stretching and reported that 90 s of stretching resulted in a decrease in passive torque \cite{magnusson1996viscoelastic}. Mizuno et al. investigated how the stiffness of muscle-tendon units changed before and after 0, 15, 30, 60, and 90 min of stretching and reported that the stiffness of the muscle-tendon unit decreased significantly immediately after stretching compared to before the 60-s stretch \cite{mizuno2013viscoelasticity}. Reisman et al. investigated the changes in muscle properties during the stretching of the plantar flexor muscles and showed that the passive torque of the muscle increased and passive tension decreased with stretching \cite{reisman2009changes}. Next, we present previous studies that investigated changes in muscle properties owing to exercise. Murayama et al. investigated whether elbow flexor stiffness changed after exercise and reported that elbow flexor stiffness increased significantly \cite{murayama2000changes}. Yanagisawa et al. showed that muscle hardness increased immediately after exercise and decreased 30 min later \cite{yanagisawa2011evaluation}. By contrast, Niitsu et al. showed that muscle hardness increased after exercise, peaked on the second day after exercise, continued to increase, and then decreased \cite{niitsu2011muscle}. Hirono et al. also investigated changes in medial gastrocnemius muscle hardness during an athletic-team training camp and reported that muscle hardness increased significantly on days 3, 4, and 5 of the training camp, and the joint range of motion decreased significantly on days 4 and 5 \cite{hirono2016changes}. Finally, we present previous studies that investigated the changes in muscle status during stretching. Ryan et al. measured joint angles during repeated stretching of the right plantar flexor muscles at constant torque. During 30 s of low-torque stretching, the right-ankle angle was shown to increase, which was considered as viscoelastic creep \cite{ryan2010viscoelastic}. Gajdosik et al. held the ankle at three dorsiflexion angles set at 100 \%, 90 \%, and 80 \% of the maximum dorsiflexion force for 60 s and measured the torque generated \cite{gajdosik2006dynamic}. The results showed that the ankle torque decreased, which was attributed to stress relaxation due to viscoelasticity. 

Thus, previous studies reported that stretching changes muscle properties and decreases hardness \cite{magnusson1996viscoelastic, mizuno2013viscoelasticity, reisman2009changes}. However, exercise has also been reported to change muscle properties and stiffness \cite{murayama2000changes, yanagisawa2011evaluation, niitsu2011muscle, hirono2016changes}. Furthermore, studies that investigated muscle changes during stretching reported that a phenomenon related to muscle viscoelasticity occurred \cite{ryan2010viscoelastic, gajdosik2006dynamic}. Most previous studies measured muscle properties before and after stretching. Moreover, many of these studies did not measure muscle properties directly, but examined changes in joint range of motion or passive torque as indirect changes in muscle properties. Few of these previous studies measured muscles continuously and examined how muscle properties change during stretching. The duration of stretching varied among studies, and a consistent view of the minimum stretching time required to achieve a stretching effect was not obtained.

Therefore, the objective of this study is to investigate how the viscoelasticity of the muscle changes during static stretching by measuring the state of the muscle during stretching using continuous time-series data. By examining these continuous time-series data, we also investigate the time required for the effects of stretching to become apparent. Furthermore, by measuring the changes in muscle viscoelasticity during static stretching before and after a short-term exercise load of running on a treadmill, we examine the effects of short-term exercise load on the changes in viscoelasticity during static stretching.

\section{\label{methods} Methods}

This study was approved by the Ethical Committee for Human Studies of the Graduate School of Engineering Science, Osaka University, Japan. All experiments were performed in accordance with the approved design. Before the experiment, all the participants signed an informed-consent form. The participants included in the figures gave informed consent for the publication of their images online.

In this study, the changes in muscle viscoelasticity were continuously captured by measuring the muscle reaction force during static stretching. In this study, we used a device that applied a force from the muscle during stretching and measured the reaction force. The device was attached to the participants, and time-series data of the reaction force during stretching were obtained.  Time-series data of the measured reaction force were applied to a spring-pot model to identify the parameters of the model before and after the change in viscoelasticity. The identified parameters of muscle viscoelasticity were examined for viscoelasticity changes during static stretching. To investigate the influence of short-term exercise on the results, measurements were obtained twice, before and after exercise. The participants were subjected to a 10-min treadmill run between the two measurements to provide an exercise load.

\subsection{\label{setup} Experimental setup}

The following describes the apparatus used to measure the muscle reaction force (Fig. \ref{fig:concept}). Figure \ref{fig:concept} shows the device used to measure the muscle reaction force. The device obtained the reaction force from the muscle by pushing the central projection (indenter) of the device into the muscle. The indenter was a cylinder with a diameter of 9 mm, which was in contact with the muscle and had a chamfer of R = 4.5 mm. A small compression-type load cell (UNSCR-10N, Unipulse Corporation, Japan) was attached to the base of the indenter to measure the force applied to the indenter. The base of the load cell was attached to the base of the device, which was curved to prevent the application of force to the muscle by the base part. A non-stretchable band was attached to the left and right sides of the base of the device and wrapped around the middle of the participant's thigh. As the muscle was enlarged by static stretching, the indenter of the device was pressed against the thigh muscle, and the reaction force was measured.

\subsection{\label{protocol} Protocol}

\begin{figure*}
\includegraphics[width=14cm]{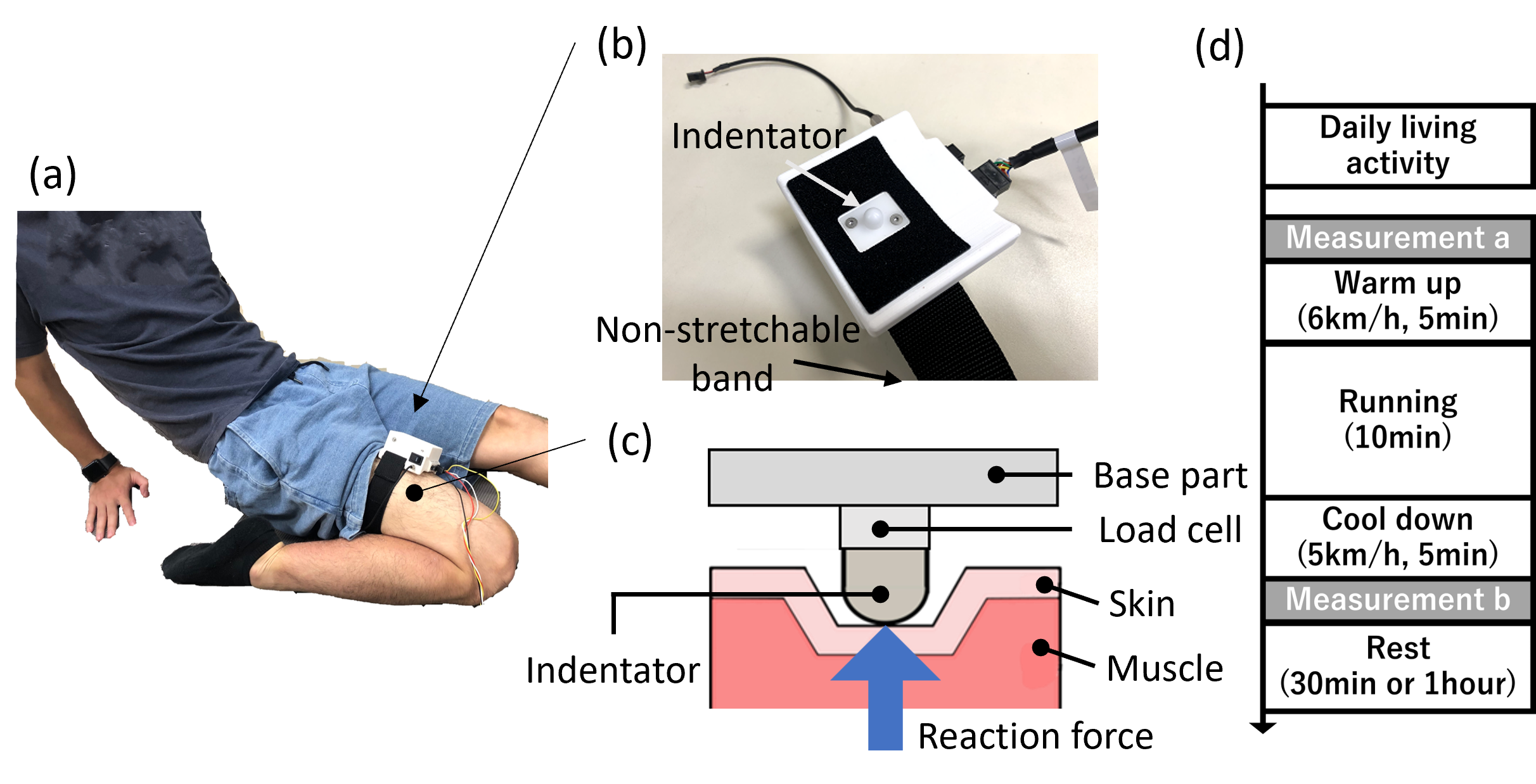}% Here is how to import EPS art
\caption{\label{fig:concept}
Experimental setup: \textbf{(a)} Picture during static stretching. A non-contractile band was used to attach the device so that the indenter was placed in the middle of the thigh. \textbf{(b)} The device obtained the reaction force from the muscle by pushing the central projection (indenter) of the device into the muscle. \textbf{(c)} A small compression-type load cell was attached to the base of the indenter to measure the force applied to the indenter. The base of this load cell was attached to the base of the device. \textbf{(d)} Experimental protocol.
}
\end{figure*}

Six healthy males with an average age of 24.1 years, daily exercise habits, and no history of injury or pain in the thigh muscles, including the rectus femoris, were tested an average of four times per participant for a total of 24 times. The experimental protocol was as follows and is shown in Fig. \ref{fig:concept}(d). 

Prior to the start of the experimental protocol, the participants were instructed to perform their normal daily routines and were not given any special exercise load. In the experimental protocol, the first experiment was performed to measure changes in muscle viscoelasticity during static stretching (measurement (a)). First, a non-contractile band was used to attach the device such that the indenter was placed in the middle of the thigh. Static stretching was then performed, during which the reaction force was measured at a sampling frequency of 1000 Hz for 90 s. The stretching period in this study was 90 s, because previous studies \cite{magnusson1996viscoelastic, magnusson1995viscoelastic} on the effects of static stretching used 90 s as the stretching period. Measurements were performed for both the left and right legs. 

A treadmill (MARCHERFT-011, Fujimori, Japan) was used to provide the participants with an exercise load via running. The treadmill speed was manually adjusted such that the participant's heart rate was maintained at the target heart rate. The heart rate during running was measured using a fitness tracker (Fitbit Inspire HR, Fitbit, USA) attached to the participant's arm. The target heart rate was calculated by the Karvonen method \cite{robergs2002surprising} using the following formula: ---$\text{target heart rate} = 0.7 \times (\text{maximum heart rate}-\text{resting heart rate}) \nonumber + \text{resting heart rate}$---. The target heart rate was defined as the value at 70 \% exercise intensity. Various methods for calculating the maximum heart rate exist \cite{she2015selection}; however, we used the simplest method: ---$\text{maximum heart rate} = 220 - \text{age}$---. The resting heart rate of each participant was measured using a fitness tracker. The participants first warmed up by running at 6.2 km/h for 5 min and then ran on the treadmill for 10 min to maintain their target heart rate, which was followed by 5 min of cool-down at 5 km/h. The exercise load and speed of the treadmill during warm-up and cool-down were based on the speeds used in previous studies in which treadmill running was performed \cite{de2003effect, wiles1992effect, morin2011sprint}.

At the end of the treadmill run, when the participant's breathing calmed, an experiment was conducted to measure the change in muscle viscoelasticity during the static stretch (measurement (b)). The method used in this experiment was the same as that used in the previous experiment before the treadmill run, and data were obtained for both the left and right feet. Thus, the experiment to measure the change in muscle viscoelasticity by static stretching was performed twice, before and after running; the first and second measurements are referred to as "measurement (a)" and "measurement (b)," respectively. Measurement (a) was the measurement of the change in muscle viscoelasticity during static stretching when each participant was not intentionally exercising. Measurement (b) shows the short-term post-exercise (treadmill running) change in muscle viscoelasticity during static stretching.

\subsection{\label{model} Viscoelastic model}

This section presents the viscoelastic model for muscle for analyzing stress-relaxation data. Muscle is a viscoelastic material \cite{fung2013biomechanics} that exhibits a mixture of elasticity, in which the strain is proportional to stress, and viscosity, in which the strain rate is proportional to stress. When a certain amount of strain is applied to a material and the material is held in place, stress is generated owing to the displacement of the molecules inside the material from their equilibrium positions if the material is elastic, whereas no stress is generated because of permanent deformation if the material is viscous. However, in a viscoelastic body, stress gradually decreases as the molecules flow internally. This is called stress relaxation and is a characteristic of viscoelastic materials \cite{fung2013biomechanics}. In general, the longer the relaxation time, the stronger is the viscous element, and the shorter the relaxation time, the stronger is the elastic element. In this study, we measured the stress relaxation that occurs when a certain amount of strain is applied to the muscle and considered the change in this trend to be a change in muscle viscoelasticity.

Biological tissues are complex viscoelastic bodies \cite{bilston2018soft}. Viscoelastic models, such as the classical viscoelastic model, which consists of springs and dashpots connected in series or parallel, require many elements and parameters; thus, they are difficult to interpret when modeling with accuracy \cite{freed2006fractional, bonfanti2020fractional}. Therefore, models using fractional calculus called "springpot models" are often used to represent the viscoelasticity of biological tissues with fewer parameters \cite{zhang2018modeling, meral2010fractional, zhang2007congruence, kiss2004viscoelastic, kobayashi2012modeling, tsukune2015automated, kobayashi2017simple, kobayashi2020non}. An infinite number of springs and dashpots converge to a single element called a springpot, which uses fractional calculus \cite{kobayashi2017simple}. The springpot model can accurately represent viscoelasticity, which is characterized by the power law, with a small number of parameters \cite{freed2006fractional, bonfanti2020fractional}. In this study, a springpot model was selected as the viscoelastic model for the muscle because it can capture characteristics with a small number of parameters to easily capture changes during static stretching.

In the spring-pot model, the relationship between stress $F(t)$ (reaction force from the muscle) and strain $\varepsilon(t)$(indentation displacement),where $t$ is time, can be written using fractional calculus as follows:

\begin{eqnarray}
F(t) = G\tau_0^r\frac{d^r \varepsilon(t)}{dt^r},
\label{eq:1}
\end{eqnarray}

where $G$ and $r$ are parameters that characterize viscoelasticity. $r$ determines the rate of stress relaxation, where a viscoelastic body is closer to an elastic element as it approaches 0 and is closer to a viscous element as it approaches 1. $G$ determines the magnitude of the force scale at a certain time $\tau_0$ during the stress relaxation. $\tau_0$ denotes the reference time scale. 

Given a constant strain $\varepsilon(t)= \varepsilon_0$ at time $t>0$, force $F(t)$ is given as follows \cite{bonfanti2020fractional}:

\begin{eqnarray}
F(t) = \frac{G \varepsilon_0}{\Gamma(1-r)} \left (\frac{t}{\tau_0} \right) ^{-r}, 
\label{eq:2}
\end{eqnarray}

where $\Gamma()$ is the gamma function. The log transformation of both sides of equation (\ref{eq:2}) yields the following equation, where $\log F(t)$ is a linear function of $\log (t)$:

\begin{eqnarray}
\log F(t) = -r \log \left(\frac{t}{\tau_0}\right) + \log \left( \frac{G \varepsilon_0}{\Gamma(1-r)} \right).
\label{eq:3}
\end{eqnarray}

Thus, in the stress-relaxation data, if the viscoelastic properties (parameters $G$ and $r$ in this model) do not change during static stretching, then plotting the relationship between time $t$ and reaction force $F(t)$ on both logarithmic plots yields a straight line. Conversely, a change in the slope or intercept of the line indicates a change in the parameters $r$ and $G$ that characterize viscoelasticity, thereby indicating a change in the viscoelasticity of the muscle. 

\subsection{\label{processing} Data processing}

This section describes the data-processing methods used to calculate the model parameters ($G$ and $r$) shown in Equation (\ref{eq:3}) from the time-series data of the muscle reaction forces. To smooth the acquired data, a moving average was taken from the obtained reaction-force time-series data with an interval of 500 [ms]. The data were then resampled so that they were equally spaced in both logarithmic plots. 

This study assumed two types of muscle states, one before and one after the change due to static stretching. Thus, we divided the muscle into two segments and calculated the parameters from the slope and intercept of the two regression lines. Specifically, segment regression \cite{muggeo2003estimating} was performed on the resampled data in both logarithmic plots. 

From the slope and intercept of the straight segment regression line, we calculated the parameters $r$ and $G$ using Equation (\ref{eq:3}). From Equation (\ref{eq:3}), the parameter $r$ is the slope of the line. Parameter $G$ can be calculated as $G=I_o \Gamma(1-r) / \varepsilon_0$, where $I_o$ is the regression-line value at $t=\tau_0$. Therefore, $\tau_0$ is related to $G$. In this study, we set $\tau_0=1$ for simplicity. This implies that in both logarithmic plots, the point at which the stress-relaxation line intersects the vertical axis of $t=1$ [s] is the standard for the magnitude of $G$. The value of $\varepsilon_0$ in Equation (\ref{eq:3}) corresponds to the amount of the indenter pushed into the muscle. Because this value cannot be measured owing to the experimental method used in this study, we set $\varepsilon_0=1$. Note that in the results and discussion that follow, we focus only on the relative changes with respect to the parameter $G$; thus, setting $\varepsilon_0=1$ has no effect on the results or discussion of this study.

The reaction force of the muscle increased instantaneously during a voluntary contraction of the participant’s muscle during measurement. To exclude voluntary contractions from the experimental data, the data were discarded based on a coefficient of determination of $R^2>0.9$ for the results obtained by segment regression. 

The Mann--Whitney U test was also performed on the parameters obtained from the data of the left and right feet at a significance level of $p<0.05$. No significant difference was observed between the results for the left and right feet in measurements (a) and (b). Therefore, for the given measurement data, the left-right foot differences were not considered.

Hereafter, the parameters from the first regression line at measurement (a) are defined as $r^a_1$ and $G^a_1$, and the parameters for the second regression line as $r^a_2$ and $G^a_2$. The boundary point between the two segments and time at which the stress-relaxation trend changed was defined as $t^a_c$. Similarly, in measurement (b), we defined the parameters of the first regression line as $r^b_1$ and $G^b_1$, those of the second regression line as $r^b_2$ and $G^b_2$, and the time of the segment boundary point as $t^b_c$. In addition, in the mixed data of measurements (a) and (b), we defined the parameters of the first regression line as $r^{ab}_1$ and $G^{ab}_1$, those of the second regression line as $r^{ab}_2$ and $G^{ab}_2$, and the time of the segment boundary as $t^{ab}_c$.

\subsection{\label{statistical} Statistical analysis}

The Wilcoxon signed-rank test \cite{woolson2007wilcoxon} was performed on the viscoelastic-model parameters $r^a_1$ and $r^a_2$ obtained from measurement (a) at a significance level of $p<0.05$. Similarly, for measurement (b), the Wilcoxon signed-rank test was conducted for $r^b_1$ and $r^b_2$ at a significance level of $p<0.05$. A Wilcoxon signed-rank test was also performed for $r^{ab}_1$ and $r^{ab}_2$ by combining the results of measurements (a) and (b) at a significance level of $p<0.05$. Wilcoxon signed-rank tests were also conducted for the parameter-change points $t^a_c$ and $t^b_c$ at a significance level of $p<0.05$.

The correlation coefficients for each parameter were calculated from measurement results of (a) and (b). The parameters considered were $r_1$, $r_2$, $G_2/G_1$, $r_2-r_1$, and $r_2/r_1$. Here, each parameter and its change $r_2-r_1$, change rate $G_2/G_1$, and $r_2/r_1$ were considered as variables. In this correlation analysis, the data for measurements (a) and (b) were not separated, but all the data ($r^{ab}_1$, $G^{ab}_1$, $r^{ab}_2$, and $G^{ab}_2$) were used.

\section{\label{result} Results}

\subsection{\label{stress} Stress-relaxation test}

\begin{figure*}
\includegraphics[width=17cm]{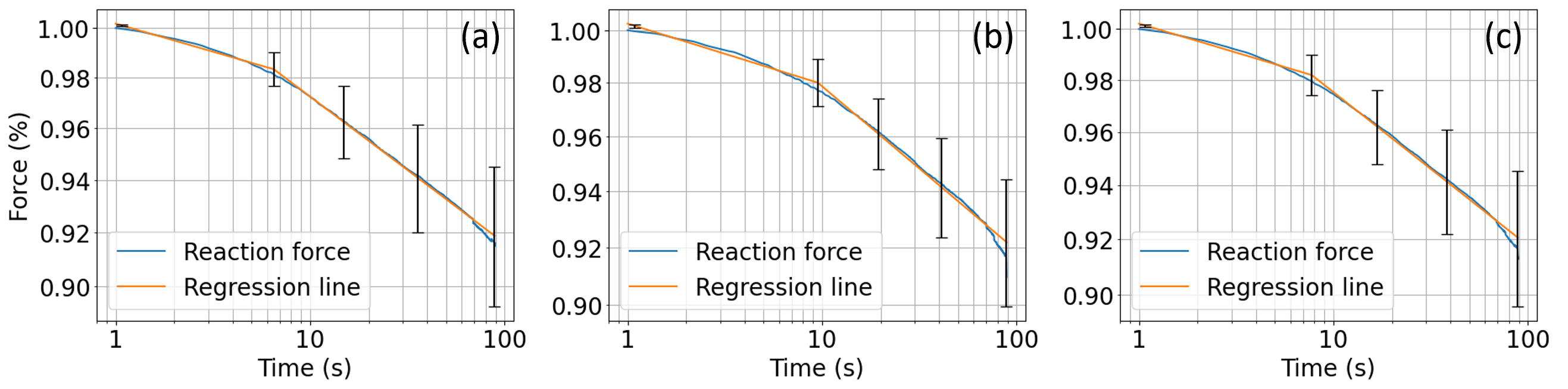}% Here is how to import EPS art
\caption{\label{fig:time} 
The time series of the averaged reaction force (stress-relaxation data) during static stretching and model of each experimental dataset. In each figure, the blue line shows the experimental result, and the orange line shows the model result (regression line). The error bars indicate standard deviation. \textbf{(a)} The data obtained in measurement (a), \textbf{(b)} the data obtained in measurement (b), \textbf{(c)} the data of both measurements (a) and (b) combined. Note that in Figs. (a), (b), and (c), the forces are expressed as a percentage with the value of the reaction force at $t=1$ [s] set to 1 to focus on relative changes. 
} %f3
\end{figure*}

Fig. \ref{fig:time}(a) shows a graph of the time series of the averaged reaction force during a stress-relaxation test (N=30) obtained in measurement (a). The data obtained from measurement (b) are shown in Fig. \ref{fig:time}(b), and the data of both measurements (a) and (b) combined (N=60) are shown in Fig. \ref{fig:time}(c). The data used to fit the model to each set of experimental data are shown in Fig. \ref{fig:time}. In this experiment, $\varepsilon_0$ was the amount by which the indenter pushed into the muscle. Because the exact value of $\varepsilon_0$ could not be determined in each experiment, the absolute value of $G$ could not be compared between each measurement. However, if the rate of change of the parameter $G_2/G_1$ is considered, the term $\varepsilon_0$ disappears, and comparisons can be made between the measurements. Note that in Figs. (a), (b), and (c), the reaction forces are expressed as percentages, with the value of the reaction force at $t=1$ [s] set to 1 to focus on the relative changes. The experimental results in Fig. \ref{fig:time} show that the stress-relaxation curve (both logarithmic plots) is a line graph comprising two segments. The stress-relaxation curve was fitted to the experimental results with high accuracy using segment regression. Moreover, the coefficient of determination for the fitted regression line is high (R2 = 0.99).

\subsection{\label{parameter} Parameter change}

\begin{figure*}
\includegraphics[width=14cm]{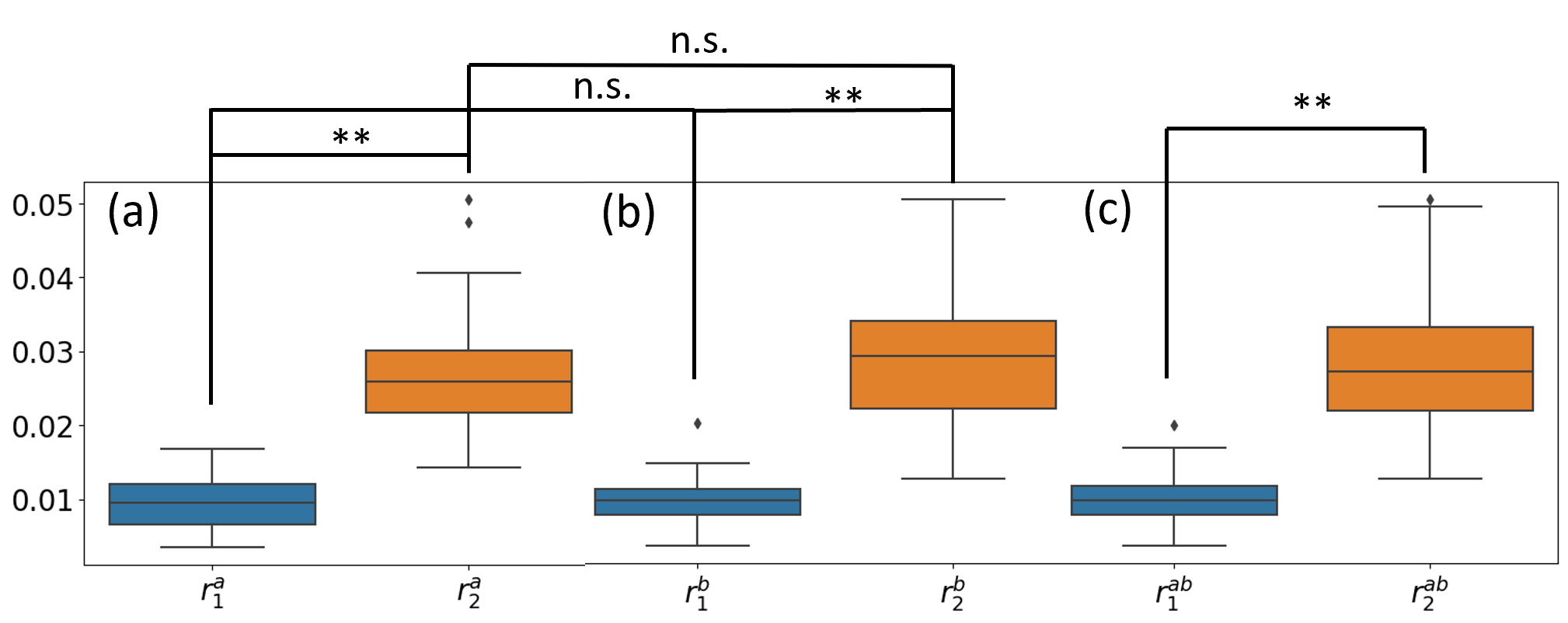}% Here is how to import EPS art
\caption{\label{fig:stats1} 
The box-and-whisker diagrams of parameter $r$ obtained for \textbf{(a)} measurement (a), \textbf{(b)} measurement (b), \textbf{(c)} the combined data of measurements (a) and (b). The statistical analysis showed that for the parameter $r$, the value of $r_2$ increased significantly with respect to $r_1$ in the data for measurement (a), measurement(b), and for both measurements (a) and (b) ($**:{\rm p}<0.01$).
} %f3
\end{figure*}

For parameters $r_1$ and $r_2$, the box-and-whisker diagrams obtained in measurement (a) are shown in Fig. \ref{fig:stats1}(a), those in measurement (b) in Fig. \ref{fig:stats1}(b), and the combined data of measurements (a) and (b) in Fig. \ref{fig:stats1}(c). The Wilcoxon signed-rank test, used as the statistical-analysis method, showed that for the parameter $r$, the value of $r_2$ increased significantly with respect to $r_1$ in the data for measurement (a), measurement(b), and for both measurements (a) and (b) ($**:{\rm p}<0.01$).

\begin{figure*}
\includegraphics[width=14cm]{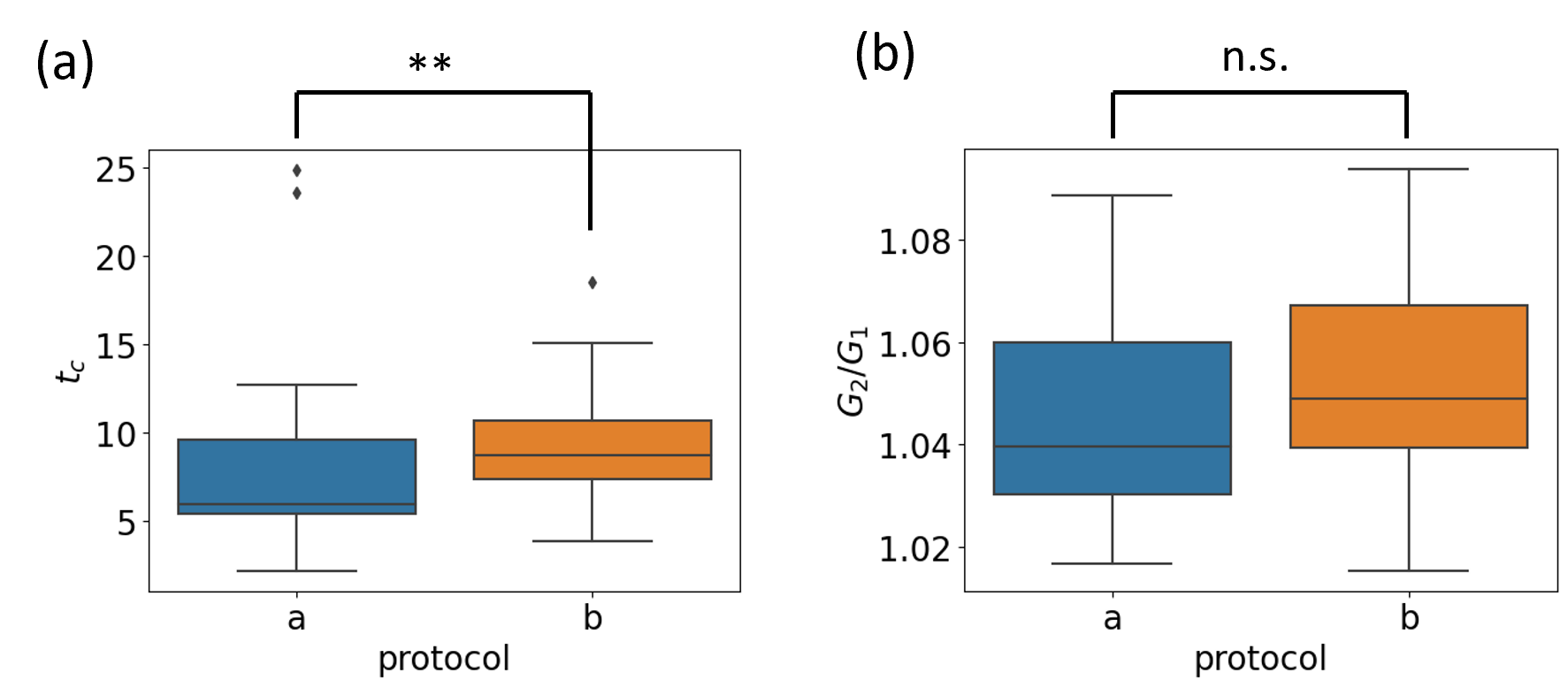}% Here is how to import EPS art
\caption{\label{fig:stats2} 
The box-and-whisker diagrams of parameter $t_c$ and $G_2/G_1$ obtained in measurements (a) and (b). \textbf{(a)} the time required until the parameter changed ($t_c$) increased significantly ($**:{\rm p}<0.01$) in measurement (b) compared to measurement (a). \textbf{(b)} No significant difference was observed for the rate of change $G_2/G_1$ between the data in measurements (a) and (b).
} %f3
\end{figure*}

Fig. \ref{fig:stats2}(a) shows a box-and-whisker plot comparing the time required until the parameter changes ($t_c$) for measurements (a) and (b). The results of the Wilcoxon signed-rank test for the results of $t_c$ in measurements (a) and (b) show that the value of $t_c$ increased significantly ($**:{\rm p}<0.01$) in measurement (b) compared to measurement (a).

Fig. \ref{fig:stats2}(b) shows a box-and-whisker plot comparing the rate of change $G_2/G_1$ of the parameter $G$ for measurements (a) and (b). Parameter $G$ cannot be compared between different measurements, because the amount of indentation into the muscle $\varepsilon_0$ cannot be measured. Therefore, the rate of change $G_2/G_1$ was examined. A Wilcoxon signed-rank test was performed on the results of $G_2/G_1$ for measurements (a) and (b), and no significant differences were observed between them ($p=0.12$). Similarly, no significant difference existed between measurements (a) and (b) for $r_1$ and $r_2$ ($r_1$: $p=0.84$; $r_2$: $p=0.40$), as shown in Fig. \ref{fig:stats1}. Therefore, in the results that follow, the results of measurement (a) are not separated from the results of measurement (b). Instead, the results are presented with respect to the parameters ($G^{ab}_2/G^{ab}_1$, $r^{ab}_1$, and $r^{ab}_2$) that combine both datasets.

\subsection{\label{parameter change} Correlation among parameters}

\begin{figure*}
\includegraphics[width=14cm]{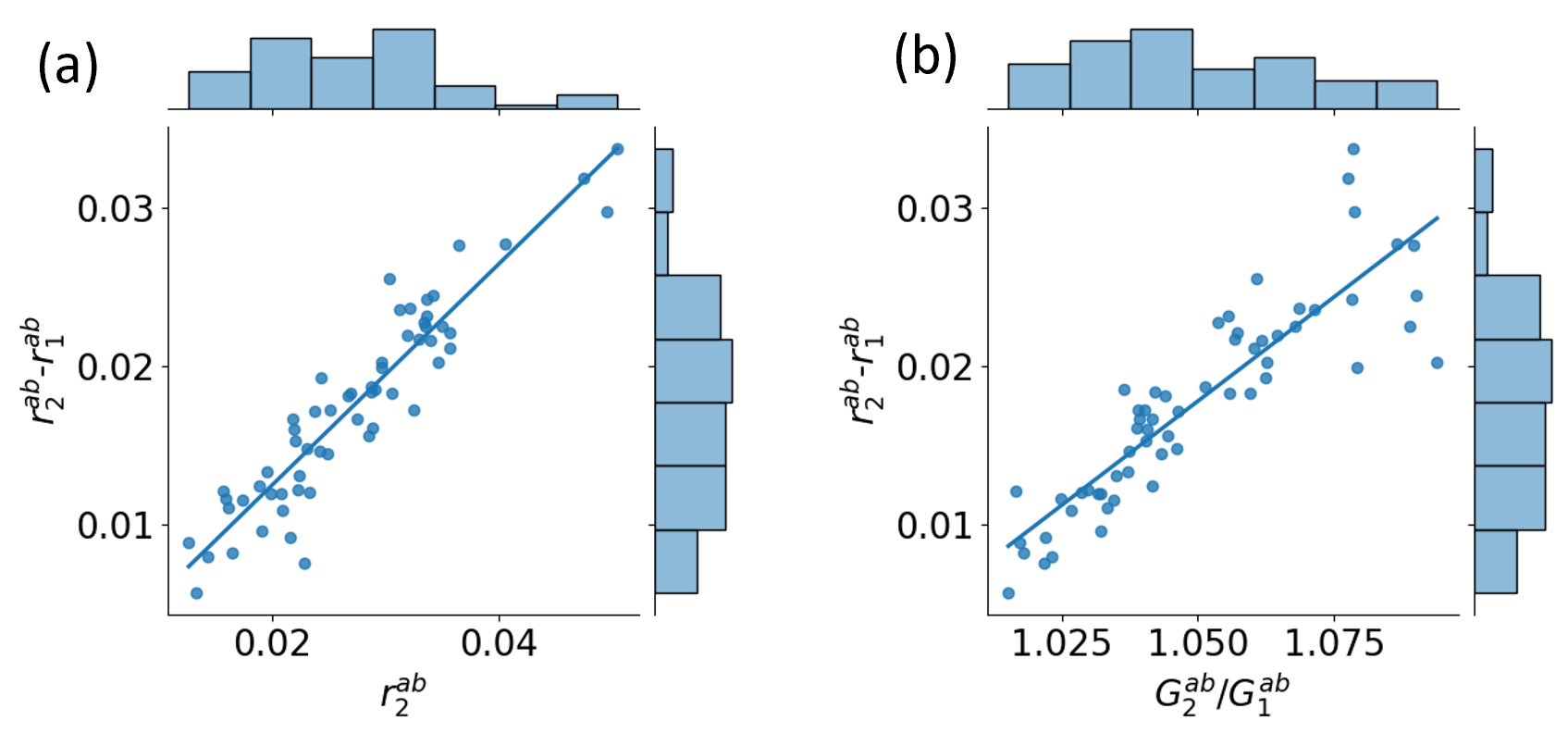}% Here is how to import EPS art
\caption{\label{fig:correlation} 
The correlation coefficients for each parameter. \textbf{(a)} shows a scatter plot and histogram of parameters $r^{ab}_2$ and $r^{ab}_2-r^{ab}_1$ with a correlation coefficient of 0.934. \textbf{(b)} shows a scatter plot and histogram of parameters $G^{ab}_2/G^{ab}_1$ and $r^{ab}_2-r^{ab}_1$ with a correlation coefficient of 0.885.
} %f3
\end{figure*}

Table \ref{param_corr} lists the correlation coefficients of each parameter. The following figures show strong correlations among the parameters. Figure \ref{fig:correlation}(a) shows a scatter plot and histogram of the parameters $r^{ab}_2$ and $r^{ab}_2-r^{ab}_1$ with a correlation coefficient of 0.934. Figure \ref{fig:correlation}(b) shows a scatter plot and histogram of the parameters $G^{ab}_2/G^{ab}_1$ and $r^{ab}_2-r^{ab}_1$ with a correlation coefficient of 0.885. Correlation coefficients between $r^{ab}_2$ and $G^{ab}_2/G^{ab}_1$ and between $r^{ab}_2$ and $r^{ab}_1$ were also high (correlation coefficients of 0.83 and 0.756, respectively).

\begin{table}[H]
\centering
\caption{Correlation coefficient between each of the parameters\label{param_corr}}
\begin{tabular}{|c|c|c|c|c|c|}
\hline
 & $r^{ab}_1$ & $r^{ab}_2$ & $r^{ab}_2$-$r^{ab}_1$ & $r^{ab}_2$/$r^{ab}_1$ & $G^{ab}_2$/$G^{ab}_1$\\ \hline
$r^{ab}_1$ & - & 0.756 & 0.472 & -0.534 & 0.426  \\ \hline
$r^{ab}_2$ & - & - & 0.934 & 0.090 & 0.830  \\ \hline
$r^{ab}_2$-$r^{ab}_1$ & - & - & - & 0.413 & 0.885  \\ \hline
$r^{ab}_2$/$r^{ab}_1$ & - & - & - & - & 0.324  \\ \hline
$G^{ab}_2$/$G^{ab}_1$ & - & - & - & - & -  \\ \hline
\end{tabular}
\end{table}

\section{\label{discussion} Discussion}

% \subsection{\label{dis_stress} Stress relaxation test}

To date, only a few studies have measured the changes in viscoelasticity during static stretching, and the change in viscoelasticity during static stretching remains unknown. The main contribution of this study is to show that viscoelasticity tends to change abruptly at a certain time during static stretching because the stress-relaxation line can be accurately approximated by segment regression on both logarithmic plots. The change in muscle viscoelasticity during static stretching was captured by applying stress relaxation to a spring-pot model (a model with fractional calculus). Considering the two states of viscoelasticity, before and after the change, the stress-relaxation curve was fitted to the spring-pot model with high accuracy via segment regression. The coefficient of determination of the fitted regression line is high (R2 = 0.99).

% \subsection{\label{dis_parameter} Change in viscoelatic parameters}

An increase in the parameter $r$ in the spring-pot model induced a faster stress relaxation. Conversely, a decrease in $r$ resulted in a slower stress relaxation. In both measurements (a) and (b) obtained in this experiment, parameter $r$ increased during static stretching. Therefore, static stretching is expected to induce faster stress relaxation. 

\begin{figure*}
\includegraphics[width=14cm]{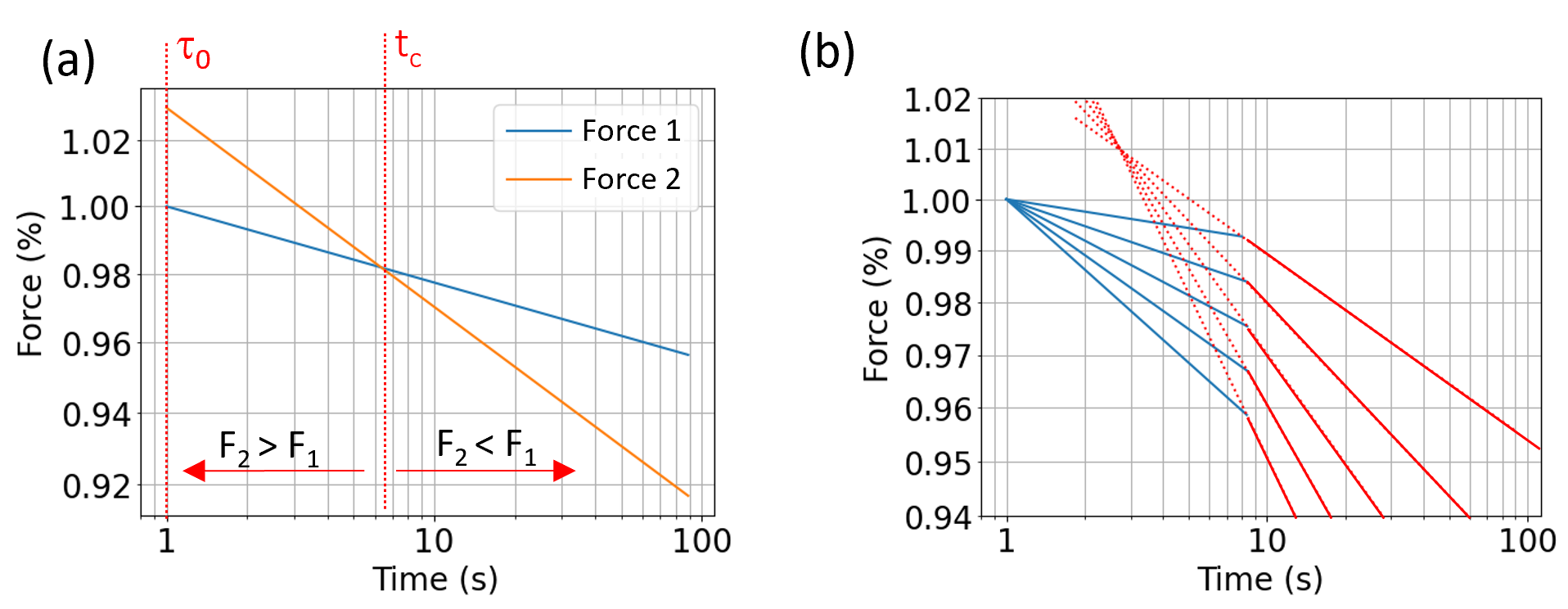}% Here is how to import EPS art
\caption{\label{fig:explain} 
\textbf{(a)} Comparison of muscle reaction forces before and after the viscoelasticity change. \textbf{(b)} Typical example of stress-relaxation test from constraints on parameters, i.e., if the muscle state at the start of static stretching (blue line) is known, the state at the end (red line) can be predicted. The line after the change in viscoelasticity due to static stretching (red line) is extended with a dotted line. The lines showing the stress relaxation after the change intersect at approximately one point. A similar tendency was observed in the research of various cells \cite{kollmannsberger2011linear}.} %f3
\end{figure*}

Measurements (a) and (b) obtained in the present experiment show that $G2/G1>1$ and $G$ increased during static stretching. However, this does not mean that static stretching increased "muscle hardness" when simply defined as "muscle hardness = reaction force/ indentation displacement." Parameter $G$ varied with $\tau_0$ in the spring-pot model. The value of $G_2/G_1$ also changed with $\tau_0$; $G_2/G_1>1$ under $\tau_0<t_c$ and $G_2/G_1<1$ under $\tau_0>t_c$. Most previous studies reported a decrease in muscle hardness after static stretching \cite{mizuno2013viscoelasticity, reisman2009changes, magnusson1996viscoelastic}. However, in these studies, to avoid the effects of stress relaxation, muscle-hardness measurements were performed at the reaction force after a sufficient time was taken to measure the muscle hardness \cite{mizuno2013viscoelasticity, reisman2009changes, magnusson1996viscoelastic}. These are equivalent to measurements under $\tau_0>t_c$ conditions, i.e., $G_2/G_1<1$ in the results of this study. In fact, comparing the regression lines before and after the change in this study, the simple reaction-force values were larger after the change before $t_c$, the time before the viscoelasticity changes, as shown in Fig. \ref{fig:explain}(a). However, after $t_c$, the value after the change decreased. Therefore, after a time lapse of $t_c$, the simple muscle hardness decreased. Thus, the results of the present study do not conflict with those of previous studies that reported a decrease in muscle hardness after static stretching.

% \subsection{\label{dis_time} Required time to change muscle viscoelasticity}

This study defined the specific number of seconds required to change muscle characteristics during static stretching. The experimental results also showed that the time required for the viscoelasticity of the muscle to change due to static stretching significantly increased after treadmill running. In measurement (a), i.e., when the participants performed only daily movements before the measurement, the change in the viscoelasticity of the muscle owing to static stretching was observed after 6.5 s. Measurement (b), obtained after short-term exercise, required approximately 9.5 s. This suggests that the time required to extend the muscle by static stretching is longer than normal for changing the characteristics of the muscle stiffened by exercise. Note that in this study, prior to the start of the experimental protocol (before measurement (a) was performed), participants were instructed to go about their normal routine and were not given any specific exercise load. The results of measurements (a) and (b) suggest that static stretching for approximately 10 s was effective. The results of this study are consistent with those of previous studies. For example, according to Bandy et al., static stretching of the hamstrings for 30 s is necessary to increase the range of motion of the hip joint \cite{bandy1997effect}. They also reported that increasing the time from 30 to 60 s did not improve flexibility. In addition, Borms et al. reported that 10 s of static stretching was sufficient to improve the flexibility of the femoral neck, and no significant differences were observed between groups that performed 10, 20, or 30 s of static stretching \cite{borms1987optimal}.

% \subsection{\label{dis_col} Parameter colleration}

Based on the correlations between the measured viscoelastic parameters shown in Fig. \ref{fig:correlation}, the correlation between $r_1$ and $r_2-r_1$ is high, suggesting that if the muscle state $r_1$ at the start of static stretching is known, the state $r_2$ at the end can also be predicted. This suggests that static stretching does not bring the muscle state back to some baseline, but rather changes it relative to the initial state. In addition, the correlation between $G_1/G_2$ and $r_2-r_1$ was high. The results suggest certain constraints on the parameters when viscoelasticity changes. By using this constraint relationship and considering it with respect to the pattern of stress relaxation during stretching, once the state at the beginning of the static stretch is known, the final state can be estimated using the correlations. The process of this estimation is shown in the Appendix \ref{app:constrain}. Figure \ref{fig:explain}(b) shows the stress-relaxation curve obtained using this constraint relationship.

In Fig. \ref{fig:explain}(b), the line after the change in viscoelasticity owing to static stretching (red line) is extended with a dotted line. Figure \ref{fig:explain}(b) shows that the lines representing stress relaxation after the change intersect at approximately one point. In a study measuring the viscoelasticity of cells \cite{kollmannsberger2011linear}, a tendency for lines to intersect at a point similar to the present result was reported and is discussed below. Kollmannsberger et al. compared the creep properties of various cells and reported that, when their creep test results were extended, straight lines intersected at a single point \cite{kollmannsberger2011linear}, which is similar to the results of this study. They reported that the cells may not be able to change their elastic and viscus properties independently. They further reported that the ratio of elastic to viscus properties must be changed when changing the stiffness. Moreover, to raise the stiffness, the cell must become more solid (i.e., the parameter $r$ must be smaller), and to lower the stiffness, it must become more fluid (i.e., the parameter $r$ must be increased). The results in Fig. \ref{fig:explain}(b) exhibit the same mechanism in the viscoelasticity of cells \cite{kollmannsberger2011linear} with respect to muscle properties after the viscoelasticity changes. In other words, some mechanisms change the ratio of elasticity to viscosity when the muscle stiffness changes, and the results of this study suggest that the viscoelasticity of the muscle may change according to these mechanisms. Here, the viscoelasticity of cells \cite{kollmannsberger2011linear} was based on tests on cells with constant normal properties and not on changes in viscoelastic properties. Thus, this study suggests that the above mechanisms may also be present when muscle properties change.

The main limitation of this study is that the indentation displacement $\varepsilon_0$ cannot be measured and is assumed as $\varepsilon_0=1$. This was unavoidable because of the nature of the experiment, in which the reaction force was measured using an indenter pushing against the thigh muscle as the muscle was enlarged by static stretching. Therefore, the overall force magnitude of the stress-relaxation data was not meaningful, and the force was expressed as a percentage, with the value at 1 s after the start of static stretching being 1. In addition, the absolute values of $G_1$ and $G_2$ are not meaningful, and only the relative values of $G_2/G_1$ are considered. Moreover, the magnitude of $G$ cannot be determined if it is correlated with other parameters. However, the inability to measure $\varepsilon_0$ does not contribute to the findings of this study.

%\section*{data availability}
%The datasets generated in this study are available from the corresponding author upon reasonable request. The data are not publicly available to protect the privacy of the participants.

\begin{acknowledgments}
This study was supported in part by Grant-in-Aid for Scientific Research from the Ministry of Education, Culture, Sports, Science and Technology (MEXT) (19H02112, 19K22878, and 23K18479), Japan. This study was supported in part by JST, PRESTO (Grant Number JPMJPR23P2), Japan.
\end{acknowledgments}

\appendix{

\section{\label{app:constrain} Parameter constraints}

The constrained relationships between $r_1$ and $r_2-r_1$ and between $G_1/G_2$ and $r_2-r_1$ are discussed below with respect to the stress–relaxation pattern during stretching. The relationship between $G_1/G_2$ and $r_2-r_1$ is obtained using the least-squares method, as follows:

\begin{eqnarray}
\frac{G_2}{G_1} = 2.98 \times(r_2-r_1) + 0.997. \label{expr reg1}
\label{eq:a1}
\end{eqnarray}

Using the least-squares method for $r_1$ and $r_2$, we obtain the following equation:

\begin{eqnarray}
r_2 = 1.86 r_1 + 0.009 \label{expr reg2}%\\ (r_2-r_1 = 0.693 r_2 - 0.0013).
\label{eq:a2}
\end{eqnarray}

In this study, the absolute value of $G$ cannot be considered because the displacement by which the indenter pushed into the muscle $\varepsilon_0$ is unknown. Therefore, $G_1 = 1$ is assumed. We assume that $G_1 = 1$ is equivalent to expressing the force as a percentage, with the value at 1 s after the start of the static stretch being 1. Assuming $G_1=1$, determining any $r_1$ yields $G_2$ and $r_2$ from Equations (\ref{eq:a1}) and (\ref{eq:a2}), respectively. In other words, from the constraints, if the state at the beginning of the static stretch ($r_1$) is known, the final states ($G_2$ and $r_2$) can be estimated using the constraints from the correlations. We selected $r_1$ values, calculated $G_2$ and $r_2$ using equations (\ref{eq:a1}) and (\ref{eq:a2}), respectively, and illustrate the results in Fig. \ref{fig:explain}(b). 

}%end of Appendix

\bibliographystyle{unsrt}
\bibliography{Manuscript} % Produces the bibliography via BibTeX.

\end{document}